\documentclass[aps,prl,twocolumn,nofootinbib,groupedaddress]{revtex4}

\usepackage{amssymb}
\usepackage{graphicx}
\usepackage{amsmath}
\begin{document}

\title{Derivation of the Born rule from the unitarity of quantum evolution }

\author{G.B. Lesovik}
\affiliation{
	L.D.\ Landau Institute for Theoretical Physics, RAS,
	142432 Chernogolovka, Russia
}

\date{\today}

\begin{abstract} In order to make the quantum mechanics a closed theory one has to derive the Born rule from the first principles, like the Schroedinger equation (SEq), rather than postulate it. The Born rule was in certain sense derived in several articles, e.g. in \cite{Deutsch} and \cite{Zurek}.
In this work some arguments of previous authors are simplified and made more ``physical''. It is shown how to derive the Born rule using the conservation of quantum state norm $\langle\Psi|\Psi\rangle$ that is the unitary evolution property determined by the SEq. It is this property that makes the probability equal to the square of the amplitude modulus. 
Possible modification of the Born rule is briefly discussed in the case when the evolution of the system is described by a nonlinear  Schroedinger equation.
We also present arguments in the spirit of the Many-World Interpretation to explain the origin of probabilistic behavior. Simply speaking, the randomness appears as a result of representing the wave function by using a detector of discrete nature that is found only in one state at a time, out of two or more possible states. 
\end{abstract}

\maketitle

\section{Introduction}
The need for interpretation of quantum mechanics was recognized because, in contrast to other theories, to describe the results of measurements, the elements like the Born rule \cite{Born} or von Neumann projection postulate
\cite{vNeumann} had to be added to the theory while these elements did not follow from the theory itself. By the theory we mean here an <equation of motion>, like the Schroedinger equation (SEq), etc. Nevertheless, during the recent years, the situation has considerably changed. There is a growing consensus that the unitary evolution of quantum system combined with the detector properties can make it possible to completely describe the process of detection. The Born rule was in some way derived in \cite{Deutsch} and \cite{Zurek}
 in the spirit of Many-World Interpretation (MWI) \cite{Everett}, \cite{deWitt:73}.
We simplify the arguments of previous authors, and show how to derive the Born rule using the fact that the quantum state norm $\langle \Psi|\Psi \rangle $ is conserved during the unitary evolution. It is this property that makes the probability equal to the amplitude modulus squared.

\section{Derivation of the Born rule}
So we are going to construct a probabilistic theory based on the Schroedinger equation. 
Let us first consider such simple 
system as a particle in a double well potential, see Fig.1.
\begin{figure}[tb]
  \centering
 \includegraphics[width=7cm]{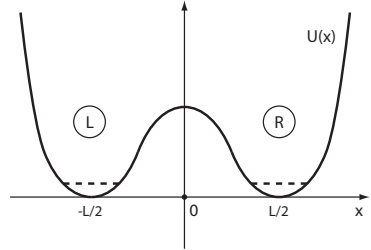}
 \caption{Particle in a double well potential. }
 \label{2well}
\end{figure}
Suppose the particle position is measured to determine if the particle is located in the left well ($x<0$) or in the right well ($x>0$). We postulate at the moment that the results of such measurement are random (the origin of probabilistic behavior will be explained later in the spirit of MWI). Now we will determine how the measurement outcome probability depends on the wave function (WF) $\Psi(x)$. 
Let the particle be in the ground state. Then its WF $\Psi $ is symmetric and can be written (we do not normalize it) as in \cite{LL3}
\begin{equation}
\Psi = 
\phi(x+L/2) +\phi(x-L/2) 
\end{equation}
Here $\phi(x) $ is the WF of the ground state in an isolated dot (well), $L$ is the distance between the dots.  
Due to the symmetry of the problem, it is natural to suppose that the probabilities $P_{L(R)}$ that the particle is in one of the dots are the same and equal to one half: 
\begin{equation}
P_{L}= P_{R}=1/2 
\end{equation}
For the time being, this can be considered as an independent {\it symmetry postulate}, but this postulate is at least much more primitive and natural than the Born rule. Below we present arguments explaining why the probabilistic description appears at all, and how this postulate can be derived. Based on the similar symmetry argument, the probability to find the particle, that is in the symmetric state, in one of the $n$ identical dots (see Fig 2)
\begin{figure}[tb]
  \centering
 \includegraphics[width=7cm]{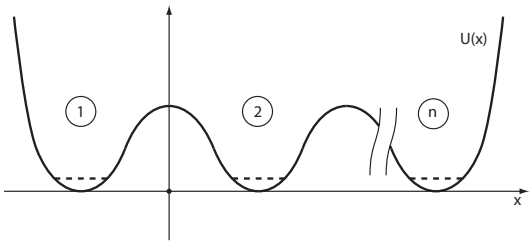}
 \caption{The probability to find a particle, which is in the symmetric state, in one of the $n$ identical dots is $1/n$.}
 \label{Nwell}
\end{figure}
 is expected to be
$1/n$:
\begin{equation}
P_{j}= 1/n, 
\end{equation}
where $j$ is the number of a particular dot. 

This expression is compatible with the Born rule that can be written using the integrals over the dot regions $\Omega_j$
\begin{equation}
P_{j}= \int_{\Omega_j} dx|\Psi_j|^2/(\sum_j \int_{\Omega_j} dx|\Psi_j|^2 )
\end{equation}
Note, nevertheless, that for any other power $m$ ($m\neq 2$) similar expression 
\begin{equation}
P_{j}= \int_{\Omega_j} dx |\Psi_j|^m/(\sum_j \int dx |\Psi_j|^m)=1/n 
\end{equation}
will also produce equal probabilities. Therefore, from the symmetry alone it is impossible to obtain the Born rule.

Let us now return to the double dot scheme and consider a more complicated case when the amplitudes in the two dots are different (we do not consider different phases yet.) 
Suppose the right dot amplitude $A_R =n$ where
$n$ is integer, and the left dot amplitude is still $A_L=1.$ Then
\begin{equation}
\Psi = 
\phi(x+L/2) + n\phi(x-L/2) 
\end{equation}
Now, the key element of derivation is the fact that if the barrier between the dots is high enough to prevent any tunneling of the wave function between the dots then we can perform the following thought experiment. 

{\it We can introduce an arbitrary deformation of the right dot that will result in no change of probability that the particle is in the left dot because there is no tunneling through the barrier.} This can be considered as the most important Statement, which is 
discussed below using two different approaches. 

First, this statement is natural from the classical point of view because the potential well deformation is carried out by applying classical field only to the right region that is isolated from the left region by a high barrier. Such manipulation with the right dot cannot change the chance of finding the particle in the left dot because of its isolation and, therefore, cannot change the probability that the particle is in the right region. So, one may conclude that the above Statement is valid due to the correspondence principle. 

Second, the above Statement could also be derived from the conservation of total probability. If we accept that the total probability remains unity, we may then note that the WF in the left dot, according to SEq, does not change during the manipulation in the right dot because of the high barrier between the dots. Then, for any dependence of the probability on the WF, the probability will remain constant in the left dot (neglecting the time dependent phase) and, therefore, it will remain constant in the right dot, since $P_R=1-P_L. $       

While keeping in mind this rule, let us perform the following thought experiment: let us deform the right dot in such a way as to produce an extra dot, see Fig 3, 
\begin{figure}[tb]
  \centering
 \includegraphics[width=7cm]{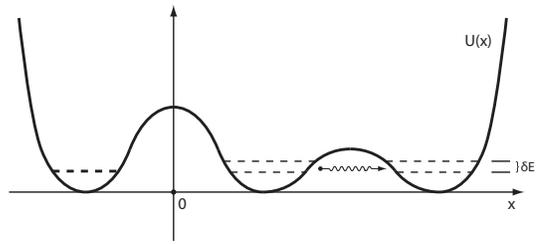}
 \caption{Let the right dot in a double well potential be extended to the right to produce an extra (third) dot. Adjusting the energy levels into resonances, one can control the penetration of the wave function into the new dot. According to the SEq, the amplitude in the new dot $A$ grows with the time $t$ as $A(t)=A(0)\sin [\delta E t /2\hbar ]$ , where $\delta E$ is the level splitting in the last pair of the dots.}
 \label{3well}
\end{figure}
so that there are three dots altogether. By adjusting the barrier between the two dots in the right region one can control the penetration of the wave function into the new dot. According to the SEq, the amplitude in the new dot $A$ changes with time $t$ as $A(t)=A(0)\sin [\delta E t /2\hbar ]$ , where $\delta E$ is the level splitting in the two dots. Similarly, we can add more dots to the right, and allow the wave function to tunnel into the new dots. Let us now ask a question, how many dots do we need in the right region in order to obtain the amplitude in all of these dots equal to unity $A_j=1?$ 
An obvious but very important answer is: 

the number of dots in the right region must be equal to $n^2$. 

This conclusion follows from the unitarity of evolution, which is determined by the SEq as the conservation of the $ \int dx |\Psi|^2$ value. In our case this integral is conserved independently in the left region $x<0$ and in the right region $x>0$ due to the negligible tunneling through the high barrier between the regions. Therefore, the above conclusion (combined with converting the event space into the space of equiprobable elementary events) is the key to the problem, so that the rest of the solution is quite straightforward.  Since in all the regions we get $1+n^2$ equivalent dots, from the symmetry of wave function configuration in all the dots one obtains the probability $P_j=1/(1+n^2)$ that the particle is in a particular dot. Therefore, total probability $P_{R}$  that the particle is in the right region (for $x>0$) (we assume here the probability additivity in the $x$-space) is 
\begin{equation}
P_{R}= n^2/(1+n^2) 
\end{equation}
which after conventional normalization to unity $\int dx |\Psi|^2=1$ yields
\begin{equation}
P_{R}= |\Psi_R|^2,
\end{equation}
where the power $m=2$ is now fixed by virtue of the fact that the integral $ \int dx |\Psi|^2$ is conserved due to the unitarity of the WF evolution. 

The above reasoning can be obviously extended to the case when ratio $A_L/A_R $ is some rational or even a square root of rational. Indeed, if $A_L/A_R=\sqrt{n/m } $ then, by changing the norm for convenience, one may obtain $A_L=\sqrt{n } $ $A_R=\sqrt{m } $, and then add $n-1$ dots to the left and $m$ dots to the right. Then the probabilities are 
\begin{equation}
P_{L}= n/(m+n) ;  P_{R}= m/(m+n) 
\end{equation}
Thus we have derived (4), i.e. the Born rule. 

The fact that we can consider only rational numbers is not a serious restriction because any real number can be approximated by a proper rational with the desired accuracy. Besides, the probability is always measured as the ratio of the numbers of outcomes, so it is always a rational number.

Similar considerations can be applied to any quantum system, although converting the situation into a set of equiprobable events can be more complicated. For a 3d-coordinate WF one can subdivide the space into small boxes and carry out manipulations with each box similar to what we have done with the dots above. 

For a WF in $k$-space one has to consider a measurement, which splits the WF into components with different $k$-vectors, like e.g. in a spectrometer. In this case all the components will be spatially separated, and with the resulting coordinate WF one may carry out all the manipulations we have done with the dots above. The part of the WF with the norm in $k$-space $\int |F(k)|^2dk $ will be transformed into coordinate WF with the norm $\int |\Psi (x) |^2 dx =\int |F(k)|^2dk/2\pi .$

For a many-body WF, the manipulation on each subset of the events requires introducing potentials that act selectively on each particle, as well as many-body interaction potentials that should have a nontrivial  dependence on the coordinates. Consideration of a  many-body WF also allows one to derive the Born rule for a mixed state. Similar generalization can also be done in the presence of dissipation, e.g. due to interaction with photons or phonons.
 
Let us now discuss possible dependence of probabilities on WF phases. First, the fact that the integral over modulus squared $ |\Psi|^2$ is conserved, and that any SEq solution is invariant under multiplying the WF by a universal factor $ \exp\{i\phi \} $ (independent of time and coordinates), already  suggests that the WF phase will not influence the probabilities. Second, introducing adiabatic  pulses of potential in one of the dots, one can adjust the phase to zero.

\section{The origin of probabilistic behavior}
One can go beyond the Copenhagen (or some other axiomatic) probabilistic interpretation and consider what happens when the measurement takes place within the Many World Interpretation that originated in Everett's works \cite{Everett}, \cite{deWitt:73}. Consider a system consisting of a particle in a ground state in a double well potential, as in the beginning of this article. After each measurement we re-prepare the state of the detector $ \phi_D^0 $ and of the system $ \Psi_s $ by allowing it to contact a low-temperature reservoir. The result of each measurement is written to a memory, which, according to the MWI, is considered within the quantum mechanics. Then we have a superposition of different states of the memory describing different histories of $N$ measurements.\footnote{It is important that the memory becomes entangled with the reservoir. As a result, after $N$ measurements the memory state is described by a diagonal density matrix with nonzero diagonal elements $1/2^N$ .} 
Then the total WF $\Psi $ will be something like
$$\Psi = \Psi_s \phi_D^0 (\sum |00011010001000111010100011\rangle $$
$$|Reservoir_{00011010001000111010100011}\rangle
+ ... ) (10). $$
Here we designate the results of the measurement as $0=L; 1=R$, while the dots denote a set of states including all the remaining configurations of units and zeros. The total number of states after $N$ measurements is $2^N$ (the above formula is written for $N=26$). 

Looking at the history of results stored in the memory, one may guess why we have to obtain randomness. In certain sense each outcome looks like a random sequence. While at the fist step it is really hard to guess whether one will obtain yes or no, with a long sequence of results one will see that the portion of zeros (or unities) is approximately the same and equal to $p=1/2.$ Actually, most of the sequences will produce such a result.\footnote{It is important to note that such arguments are correct only for the symmetric case, when the amplitudes in the dots are equal.}

In the limit of $N = \infty $ practically any sequence will produce probability $p=1/2$ in accordance with the law of large numbers. More precisely, for any small $\alpha $ and $\beta $ there is such $N$ that less than $\alpha 2^N $ sequences will produce the ratio of appearances of e.g. zero $P_{L}$ as $|P_{L} -1/2|< \beta .$ The number of sequences with a given number of zeros is determined by combinatorics and can be calculated exactly, using the binomial distribution. For large $N$ the distribution becomes Gaussian, and for that case the simple estimate from above gives $ N > \ln\{1/\alpha\}/\beta^2 .$ {\it Generally, one can say that quantum randomness stems from representing the wave function using a detector of discrete nature, which possesses only one state at a time out of two or more possible states. }

\section{Conclusion}

It is shown in this work that the Born rule can be derived from the unitarity of quantum dynamics and additional assumption that the probability is a function of quantum amplitude.\footnote{A reverse statement is usually proven in textbooks on quantum mechanics (see, e.g. \cite{LL3}) where the unitarity of quantum evolution is derived from the conservation of probability that is defined according to the Born rule.} 

This result suggests that quantum mechanics is a complete theory that provides certain methods for the description of reality while these methods can be derived from the equations of motion and some additional natural assumptions. 

This work is supported by RFBR Grant 14-02-01287.
  
The author is thankful to G.-M. Graf, A. Ilyin, L. Ioffe, G. Volovik, M. Ivanov, G. Amosov, and M. Suslov  for useful discussions.

\end{document}